# The fundamental importance of discourse in theoretical physics

Philip V. Fellman[1], Jonathan Vos Post[2] and Christine M. Carmichael[3]

[1] Southern New Hampshire University, Manchester, NH, USA

shirogitsune99@yahoo.com

[2] Computer Futures, Inc., Altadena, CA, USA

jvospost3@gmail.com,

[3] Woodbury University, Burbank, CA, USA

charmech@gmail.com

## 1   Introduction

In "The Trouble With Physics", Lee Smolin provides an admirable overview of the problems facing contemporary physics [1]. We often paraphrase Smolin's analysis by tracing the boundaries where physical theory begins to break down. Newtonian kinematics runs into trouble with three bodies as Poincaré explained in pioneering what became chaos theory, even though for two bodies we get lovely Keplerian dynamics (as Feynman gave in his original elementary exposition).[2] General Relativity runs into trouble with one body at two scales: (a) Very small bodies, length near or less than a Planck length, and totally breaking down as length approaches zero; as well as (b) the universe(s) as a whole. Likewise, Quantum Field Theory also has trouble with one body, as it interacts with its own field, in a way which requires renormalization. String Theory runs into trouble with zero bodies, as it predicts $10^{500}$ or $10^{1000}$ different vacuums, of which we don't know which one we ever had. Running throughout this body of theory (i.e., classical and modern physics) is a series of implicit assumptions about time and the nature of time. This is the result of what Julian Barbour describes as "the failure to discuss duration at a foundational level" [3] [4]. A goodly portion of the obscurity surrounding the topic of duration comes not from lack of discussion but rather from lack of a precise, modern, fundamental treatment of time. [11][12][13]

## 2   The Role of Discourse Relativity – Clocks and Time

Modern physics is a complex, specialized discipline with its own highly abstract mathematical and conceptual language, which often makes it difficult for outsiders to distinguish the differences between competing arguments regarding physical theory. Anyone who doubts this need merely read Julian Barbour's 2008 winning essay for the FqXi competition, "The Nature of Time" which won the juried competition not on the basis of a new, complex mathematical treatment, but rather on the basis of the exposition and correction of a number of fundamental errors in discourse regarding the elements of physical systems, physical state, and clocks made by both Einstein and Newton.[4]

> The theory of duration and clocks that emerges from observable differences as made explicit in (3) is very different from the view that prevailed among the great relativists in the early 20[th] century. It will suffice to consider Einstein's definition of clock in 1910:
>
> > 'By a clock we understand anything characterized by a phenomenon
> > passing periodically through identical phases so that we must assume,
> > by the principle of sufficient reason, that all that happens in a given
> > period is identical with all that happens in an arbitrary period.'
>
> I see several problems with this definition. First no system ever runs through truly identical phases, so this is an idealization that hides the true nature of time; it lacks Poincare's insistence, which I have repeated, that only the universe and all that happens in it can tell perfect time. Second, Einstein's clock cannot measure time continuously. It can only indicate that a given interval has elapsed when identical phases recur. It can say nothing about the passage of time in intervals within phases. Since the universe is the only perfect clock, it seems nothing at all can be said about the passage of time unless there is recurrence of eons under identical conditions. Even then one can only say that the eons are equally long. In contrast, the ephemeris time defined by (3) runs continuously and in no way relies on recurrence of identical phases. Finally, by relying on periodicity, Einstein's definition fails to identify the true dynamical basis of time keeping and the importance of understanding why clocks can march in step.

Similarly, Peter Lynds treatment of time which is discussed in section five, aims at correcting fundamental errors in discourse with respect to the treatment of time in both classical and quantum mechanics [12].

## 3   The Role of Discourse in Quantum Mechanics – Interpretations and Non-locality

John Bell gives a wonderful summary of the role of discourse in theoretical physics and the difficulties of coming to grips with it in his 1984 paper, "Bertlmann's Socks and the Nature of Reality"[14], beginning with Einstein's argument that:

> If one asks what, irrespective of quantum mechanics, is characteristic of the world of ideas of physics, one is first of all struck by the following: the concepts of physics relate to a real outside world…It is further characteristic of these physical objects that they are thought of as arranged in a space-time continuum. An essential aspect of this arrangement of things in physics is that they lay claim, at a certain time, to an existence independent of one another, provided these objects 'are situated in different parts of space.

This is essentially the philosophical position advocated by Ludwig Wittgenstein in the "Tractatus Logico-Philosophicus".[15] Even without the benefit of subsequent arguments, including those by Wittgenstein himself, who ultimately rejected this viewpoint, [16] Bell gives us Einstein's own qualification from quantum mechanics:

> "There seems to me no doubt that those physicists who regard the descriptive method of quantum mechanics as definitive in principle would react to this line of thought in the following way: they would drop the requirement…for the independent existence of the physical reality present in different parts of space; they would be justified in pointing out that the quantum theory nowhere makes explicit use of this requirement."

Einstein then goes on to complete the argument by explaining that he nonetheless sees no reason why this requirement would necessarily have to be abandoned and that this was why he believed quantum mechanics to be an incomplete theory. What is noteworthy in regard to Bell's raising this argument is that the contemporary discourse of the opposing school of thought is even less satisfying than the explanation provided by Einstein. Bell's argument begins with the problematical position of Bohr. "Bohr once declared when asked whether the quantum mechanical algorithm could be considered as somehow mirroring an underlying quantum reality: 'There is no quantum world. There is only an abstract quantum mechanical description. It is wrong to think that the task of physics is to find out how Nature is. Physics concerns what we can say about nature." Not only is this bad philosophical discourse, but it is bad physics. Bell illustrates the scope of Bohr's problem by noting that:

> While imagining that I understand the position of Einstein as regards EPR correlations, I have very little understanding of his principal opponent, Bohr. Yet most contemporary theorists have the impression that Bohr got the better of Einstein in the argument and are under the impression that they themselves share Bohr's views. As an indication of those views, I quote a passage from his reply to Einstein, Podolsky and Rosen. It is a passage which Bohr himself seems to have regarded as definitive, quoting it himself when summing up much later. Einstein, Podolsky and Rosen had assumed that '…if, without in any way disturbing a system, we can predict with certainty the value of a physical quantity, then there exists an element of physical reality corresponding to this physical quantity'. Bohr replied: '…the wording of the above mentioned criterion…contains an ambiguity as regards the meaning of the expression 'without in any way disturbing a system'. Of course there is in a case like that just considered no question of mechanical disturbance of the system under investigation during the last critical stages of the measuring procedure. But even at this stage there is essentially the question of *an influence on the very conditions which define the possible types of predictions regarding the future behavior of the system*…their argumentation does not justify their conclusion that quantum mechanical description is essentially incomplete…this description may be characterized as the rational utilization of all possibilities of unambiguous interpretation of measurements, compatible with the finite and uncontrollable action between the objects and the measuring instruments in the field of quantum theory'

Finally, Bell explains why this is both bad philosophy and bad physics:

> Indeed, I have very little idea what this means. I do not understand in what sense the word 'mechanical' is used, in characterizing the disturbances which Bohr does not contemplate, as distinct from those which he does. I do not know what the italicized passage means – "an influence on the very conditions…" Could it mean just that different experiments on the first system give different kinds of information about the second? But this was just one of the main points of EPR, who observed that one could learn *either* the position *or* the momentum of the second system. And then I do not understand the final reference to 'uncontrollable interactions between measuring instruments and objects.' It seems just to ignore the essential point of EPR that in the absence of action at a distance, only the first system could be supposedly disturbed by the first measurement and yet definite predictions become possible for the second system. Is Bohr just rejecting the premise - "no action at a distance" – rather than refuting the argument?

## 4 The Role of Discourse in Quantum Cosmology - Maximality

As to errors of the fixable kind, C.J.S. Clarke [17] provides us with an excellent, if unintentional example. In discussing the Hawking-Penrose theorems, Clarke summarizes the conditions for a space-time to be necessarily causally geodesically incomplete (what Barbour would call a case of "sufficient reason"). Where Clarke runs into expositional difficulties is with respect to the causality condition (i.e., that "M contains no closed timelike curves"). Up to this point, he successfully demonstrates that the null geodesics from some point are eventually focused or that the null geodesics from some 2-surface are all converging and that for every non space-like vector K we have $R_{ab}K^aK^b > 0$ (3) and that every non space-like geodesic with tangent vector K contains a point at which $K_{[aRb]}cd_{[eKf]}K^cK^d \neq 0$. (p. 11) The only remaining condition is the causality condition, which he deals with by arguing that "that the part of the universe that we can see is close to a Friedmann model, which does satisfy condition (2), even though the universe as a whole may not do so." And indeed, this condition appears to be satisfied by the Yorke scaling and gravitational degrees of freedom arguments of Barbour et al. cited earlier. However, Clarke also argues that because the Friedmann solutions contain a Cauchy surface, at least a part of our universe must be (geodesically) incomplete. At this point Clarke makes a fundamental discourse error, but one which is easily fixed, and once fixed, allows his theoretical exposition to continue uninterrupted (p. 12):

> "At this stage of the argument, one's conclusions depend on what attitude is taken to maximality. One possibility might be that D(S) is a maximal Cauchy development (the largest possible space-time that can be determined by Einstein's equations on the basis of data on S). And that even if the space-time is not singular, it comes to an end because there exists no determinate equation of evolution that can fix one continuation rather than an another. On the other hand, if one accepts that space-time must be maximal then there are two other possibilities…"

The discourse error here comes from the unfortunate phrasing, "there exists no determinate equation of evolution that can fix one continuation rather than an another". If we think about Bell's characterization of the discourse problem between Einstein and Bohr and their disagreement on the subject matter of quantum mechanics, Clarke's error is easy to spot. The equations of evolution (or if one prefers, determinate equations of evolution) are a description of (to use Einstein's phraseology) "an element of physical reality corresponding to this physical quantity" and it is this physical reality which determines both the equations of evolution and the processes to which they correspond. The lack of an equation cannot possibly be the determination or the cause of the lack of a determinate evolution of space-time. This is a Bohr-like error in the confounding of the language of physics with the subject of physics. *The subject of physics is physical systems, not the language of the description of physical systems.* To believe otherwise is to imagine that equations of state have a determinative role in the experimental process. [14] Clarke's error is easily fixed, simply by removing the offending word "equation" in which case his argument becomes completely sensible as "…even if the space-time is not singular, it comes to an end because there exists no determinate evolution that can fix one continuation rather than another." Unfortunately, not all discourse errors in physics are so easily amenable to correction.

## 5 Common Sense, Time and Mechanics

Our notions of time, instants and the flow of time most frequently enter our thinking and discourse as the result of our ordinary experience. As Lynds explains this is primarily the neurobiological function of our perception of intervals of relatively short duration as "present" moments in a continuous or "flowing" stream of time [18]. While this may be subjectively satisfying and almost universally experienced it introduces fundamental errors into the discourse of physics. Unfortunately, when scientific research attempts to refute such commonly experienced and widely held notions, like those of temporal instants or instantaneous transformation, the exposition is often met with knee-jerk criticism. The Lynds paper in Foundations of Physics Letters [12] was initially rejected by many readers as a case of not understanding differential calculus. On the contrary, the differential calculus is an excellent and useful abstraction, working very much in the same way that "classes of colors" are described as a "logical fiction" in the Problems of Philosophy.[23] In short, it's not that instantaneous transformations are not useful as an approximation of the behavior of physical systems, but at some more fundamental level it becomes important to understand that in the limiting case they are mere approximations and that, in fact, "time does not flow" nor is there any quantizable or otherwise dimensionless, static instant in time. [12][13]

> Time enters mechanics as a measure of interval, relative to the clock completing the measurement. Conversely, although it is generally not realized, in all cases a time value indicates an interval of time, rather than a precise static instant in time at which the relative position of a body in relative motion or a specific physical magnitude would theoretically be precisely determined. For example, if two separate events are measured to take place at either 1 hour or 10.00 seconds, these two values indicate the events occurred during the time intervals of 1 and 1.99999…hours and 10.00 and 10.0099999…seconds, respectively. If a time measurement is made smaller and more accurate, the value comes closer to an accurate measure of an interval in time and the corresponding parameter and boundary of a specific physical magnitudes potential measurement during that interval, whether it be relative position, momentum, energy or other. Regardless of how small and accurate the value is made however, it cannot indicate a precise static instant in time at which a value would theoretically be precisely determined, because there is not a precise static instant in time underlying a dynamical physical process. If there were, all physical continuity, including motion and variation in all physical magnitudes would not be possible, as they would be frozen static at that precise instant, remaining that way. Subsequently, at no time is the relative position of a body in relative motion or a physical magnitude precisely determined, whether during a measured time interval, however small, or at a precise static instant in time, as at no time is it not constantly changing and undetermined. Thus, it is exactly due to there not being a precise static instant in time underlying a dynamical physical process, and the relative motion of body in relative motion or a physical magnitude not being precisely determined at any time, that motion and variation in physical magnitudes is possible: there is a necessary trade off of all precisely determined physical values at a time, for their continuity through time.

This simple, but very counter-intuitive conclusion has been developed in subsequent papers [20][21][22][23]. The following section explores some of these results, and one might also wish to keep in mind Julian Barbour's maxim that "had duration been properly studied in classical physics, its disappearance in the conjectured quantum universe would have appeared natural." [4]

> As a natural consequence of this, if there is not a precise static instant in time underlying a dynamical physical process, there is no physical progression or flow of time, as without a continuous and chronological progression through definite indivisible instants of time over an extended interval in time, there can be no progression. This may seem somewhat counter-intuitive, but it is exactly what is required by nature to enable time (relative interval as indicated by a clock), motion and the continuity of a physical process to be possible. Intuition also seems to suggest that if there were not a physical progression of time, the entire universe would be frozen motionless at an instant, again as though stuck on pause on a motion screen. But if the universe were frozen static at such a static instant, this would be a precise static instant of time: time would be a physical quantity. Thus, it is then due to natures very exclusion of a time as a fundamental physical quantity, that time as it is measured in physics (relative interval), and as such, motion and physical continuity are indeed possible.

It might also be argued in a more philosophical sense that a general definition of static would entitle a certain physical magnitude as being unchanging for an extended interval of time. But if this is so, how then could time itself be said to be frozen static at a precise instant if to do so also demands it must be unchanging for an extended interval of time? As a general and sensible definition this is no doubt correct, as we live in a world where indeed there is interval in time, and so for a certain physical magnitude to be static and unchanging it would naturally also have to remain so for an extended duration, however short. There is something of a paradox here however. If there were a precise static instant underlying a dynamical physical process, everything, including clocks and watches would also be frozen static and discontinuous, and as such, interval in time would not be possible either. There could be no interval in time for a certain physical magnitude to remain unchanging. Thus this general definition of static breaks down when the notion of static is applied to time itself. We are so then forced to search for a revised definition of static for this special temporal case. This is done by qualifying the use of stasis in this particular circumstance by noting static and unchanging, with static and unchanging as not being over interval, as there could be no interval and nothing could change in the first instance. At the same time however, it should also be enough just to be able to recognize and acknowledge the fault and paradox in the definition when applied to time.

This position reflects that of Barbour et al, particularly with respect to their work on the dynamics of shape and on Yorke scaling [6] and the Lichnerowicz-Yorke equation [7] and [8]. But there is a further, fundamental discourse error in general relativity caused by Minkowski's confounding of the structure of geodesics with the flow of time. In this case, the Minkowski world-line of a particle is represented as existing even when there is no relative motion or change, arguing that even when nothing happens, time still passes. As Lynds, Barbour, Smolin and others argue, (1) time must be derived from behavior within the light cone and is in this sense endogenous to the light cone and cannot be separated from the interaction of particles, forces and fields within the universe and (2) in this context it may be that time is not a proper first order variable of physical theory, but is rather emergent from these reactions. [3][4][5][6][7][8][9][10][11][12][24]. In this sense, general relativity has been largely limited in its further development precisely by poor discourse.

## 6  The Lynds-Bell Metric

In the context of defining a temporal interval rather than a static instant in time, J.S. Bell provides an interesting potential metric which allows for a more precise statement of duration while allowing for the kind of measurement error explained by Lynds [12][13]. In "Beeables for quantum theory", Bell argues with respect to dynamics that:

> For the time evolution of the state vector we retain the ordinary Schrodinger equation,
>
> $d/dt |t\rangle = iH|t\rangle$ where H is the ordinary Hamiltonian operator. (4)
>
> For the fermion number configuration we prescribe a stochastic development. In a small time interval $dt$ configuration $m$ jumps to configuration $n$ with transition probability
>
> $dtT_{nm}$, (5)    where  (6)
>
> $$T_{nm} = \frac{J_{nm}}{D}; \text{ and } \quad (7) \quad J_{nm} = \sum_{qp} 2\text{Re}\langle t|nq\rangle\langle nq|-iH|mp\rangle\langle mp|t\rangle$$
>
> and    $D_m = \sum |\langle mq|t\rangle|^2$    (8)
>
> provided $T_{nm} > 0$ if $J_{nm} \leq 0$  (9)

From (5) the evolution of a probability distribution $P_n$ over configurations $n$ is given by:

$$d/dt P_n = \sum_m (T_{nm}P_m - T_{mn}P_n) \quad (10)$$

Following Bell we do agree that the mathematical consequence of this intervallic interpretation of the Schrodinger equation (using the stochastic perturbation over a small interval as the error term) is:

$$d/dt|\langle nq|t\rangle|^2 = \sum_{mp} 2\text{Re}\langle t|nq\rangle\langle nq|-iH|mp\rangle\langle mp|t\rangle \text{ or}$$

$$d/dt D_n = \sum_m (J_{nm} = \sum_m (T_{nm}D_m - T_{mn}D_n)$$

At this point we diverge significantly from Bell's interpretation because he then uses his exposition to set up a cosmological 3-space and 1-time, Hamiltonian and initial state vector |0⟩. Our differences are two-fold. First we agree with Barbour et al. that time is actually a second order endogenous variable of the absolute cosmological manifold shape [5][6][7][8][9][10] and secondly, we have offered elsewhere our own unique solution to "the problem of specialness" [19][20][21]. We agree with Bell that this interpretation of the Schrodinger equation has unwelcome consequences for thermodynamic reversibility, some of which we have discussed previously [20][21] and others which will be more fully elucidated in a forthcoming paper.

# 7 Conclusion: Just because we all know it doesn't mean it's true

Proper discourse is difficult and complex. [29][30]. Simplifying heuristics like "Occam's razor" are often misleading and obscure rather than clarify discourse [27][28]. It is no accident that Bell titled his collected writings "Speakable and Unspeakable in Quantum Mechanics" following the conclusion of Ludwig Wittgenstein's famous essay, "On the impossibility of any future metaphysics".[1] [16] A major goal of Bell was to show how the discourse of both the Copenhagen school and the Bohmian interpretation of quantum mechanics led to inconsistent, obscure and impossible conclusions, all of which make for bad physics.

Ultimately, when we discuss "the body of theory", (i.e., classical and modern physics) we are pointing out that "discourse" promiscuously mixes language and meta-language. There is no consensus on the topology and structure of the space of all possible theories of mathematical physics, a proper subset of what Zwicky called the "Ideocosm" – the space of all possible ideas [31]. One way of looking at this is to recognize that there are *many* models of physics, as if from different switch settings (thermal, quantum and relativistic) of what was called the "Model-o-tron" in a recent discussion of Category Theory led by Jeffrey Morton.[32] Here, we argued that there is an impossible tangle in discourse if we have neither a consensus definition of "state" nor of "time" and we yet need to discuss the evolution of quantum states over time.[2]

In this context we observed that gazing at the control panel of the Model-o-tron, we see Thermal, Quantum, and Relativistic switches. The meta-model of the Model-o-tron has the first two each as binary. There are three settings for the Relativistic switch. Does this then mean that all 2 x 2 x 3 = 12 Models are known and equally valid, in an abstract sense? More subtly, the Quantum switch is in deformation meta-mode, often shown as allowing one to set Planck's Constant, with the zero setting yielding classical physics. Likewise, the Relativistic switch is sometimes portrayed as a variable C (or $1/c$) with $1/c = 0$ being Newtonian dynamics. Even more subtly, are we *sure* that Planck's constant is a real number? The two best measurements at NIST differ with statistical significance. Could $\hbar$ be a complex number, with a small non-zero imaginary component? Could it be quaternionic or octonionic? What can we really say about the topology of metrics of the manifold of settings of the Model-o-tron? Is it a complex manifold? Does it have singularities? What is its Betti number? Is it a fibration of something we already know?"[32]

Finally, in quantum cosmology, there is the issue that Barbour et al, raise, which is that "a most strange feature of general relativity and the Big Bang cosmology" is that "… in these theories, overall size is absolute, in contrast to everything else. [5][6][7][8] This is the feature of general relativity that allows the 'expansion of the universe'. In the standard model, the universe is doing two things simultaneously: it is expanding and changing its shape (it is becoming more inhomogeneous). If the universe were perfectly relational, it could only change its shape. Attractive as this idea is, it appears to be in strong conflict with the evidence from cosmology."[25][26] Our brief, foundational discussion of time, non-locality, maximality and gravity has been designed to illustrate our central point, which is that *without proper discourse, there is no proper physics*.

---

[1] Whose concluding sentence is usually translated as "That whereof one cannot speak, that thereof one must be silent".
[2] Morton's basic characterization was that the "Thermal" switch varies whether or not we're talking about thermodynamics or ordinary mechanics. The "Quantum" switch varies whether we're talking about a quantum or classical system. The "Relativistic" switch,…specifies what kind of invariance we have: Galileian for Newton's physics; Lorentzian for Special Relativity; general covariance for General Relativity.